\documentclass[12pt,twoside]{article}
\usepackage{a4wide,latexsym,graphicx,epsfig,psfrag}
\usepackage{slashed}
\usepackage{graphics}
\usepackage{epsfig}
\usepackage{amsmath}
\usepackage{amssymb}
\usepackage{array}
\usepackage{wrapfig}
\usepackage{color}

{\catcode `\@=11 \global\let\AddToReset=\@addtoreset}
\AddToReset{equation}{section}

\newcommand{\eS}{{\epsilon_S}}
\newcommand{\eT}{{\epsilon_T}}
\newcommand{\eP}{{\epsilon_P}}
\newcommand{\eR}{{\epsilon_R}}
\newcommand{\eL}{{\epsilon_L}}
\newcommand{\eLc}{{\epsilon_L^{(c)}}}
\newcommand{\eLv}{{\epsilon_L^{(v)}}}
\newcommand{\teS}{{\tilde{\epsilon}_S}}
\newcommand{\teT}{{\tilde{\epsilon}_T}}
\newcommand{\teP}{{\tilde{\epsilon}_P}}
\newcommand{\teL}{{\tilde{\epsilon}_L}}
\newcommand{\teR}{{\tilde{\epsilon}_R}}

\def\greaterthansquiggle{\raise.3ex\hbox{$>$\kern-.75em\lower1ex\hbox{$\sim$}}}
\def\lessthansquiggle{\raise.3ex\hbox{$<$\kern-.75em\lower1ex\hbox{$\sim$}}}
\newcommand{\beq}{\begin{equation}}
\newcommand{\eeq}{\end{equation}}
\newcommand{\beqa}{\begin{eqnarray}}
\newcommand{\eeqa}{\end{eqnarray}}
\newcommand{\beqan}{\begin{eqnarray*}}
\newcommand{\eeqan}{\end{eqnarray*}}
\newcommand{\ba}{\begin{array}}
\newcommand{\ea}{\end{array}}

\newcommand{\bea}{\begin{eqnarray}}
\newcommand{\eea}{\end{eqnarray}}

\def\nz{\ifmmode {I\hskip -3pt N} \else {\hbox {$I\hskip -3pt N$}}\fi}
\def\zz{\ifmmode {Z\hskip -4.8pt Z} \else
       {\hbox {$Z\hskip -4.8pt Z$}}\fi}
\def\qz{\ifmmode {Q\hskip -5.0pt\vrule height6.0pt depth 0pt
       \hskip 6pt} \else {\hbox
       {$Q\hskip -5.0pt\vrule height6.0pt depth 0pt\hskip 6pt$}}\fi}
\def\rz{\ifmmode {I\hskip -3pt R} \else {\hbox {$I\hskip -3pt R$}}\fi}
\def\cz{\ifmmode {C\hskip -4.8pt\vrule height5.8pt\hskip 6.3pt} \else
       {\hbox {$C\hskip -4.8pt\vrule height5.8pt\hskip 6.3pt$}}\fi}

\def\au{{\setbox0=\hbox{\lower1.36775ex%
\hbox{''}\kern-.05em}\dp0=.36775ex\hskip0pt\box0}}
\def\ao{{}\kern-.10em\hbox{``}}

\normalsize
\begin{document}
\begin{titlepage}
\begin{flushright}
LA-UR-12-24810\\
NPAC-12-14
\end{flushright}
\vspace{2.5cm}
\begin{center}
{\Large \bf 
%
%
Non-standard Charged Current Interactions:
\\
$ $
\\
beta decays versus the LHC
}
\\[40pt]
Vincenzo Cirigliano$^{1}$, Mart\'in Gonz\'alez-Alonso$^2$, Michael L. Graesser$^{1}$
\vspace{1cm}

${}^{1)}$ Theoretical Division, Los Alamos National Laboratory, Los Alamos, NM 87545, USA \\[10pt]
${}^{2)}$ Department of Physics, University of Wisconsin-Madison, \\
1150 University Ave., Madison, WI, 53706, USA
\vfill
{\bf Abstract} \\
\end{center}
\noindent
We discuss  low-energy and collider constraints  on 
the effective  couplings  characterizing non-standard charged current interactions.
A direct comparison of low-energy and LHC probes can be performed  within 
an effective theory framework,  when the  
new physics mediating these  interactions originates in the multi-TeV scale.  
We find that for the  effective couplings involving right-handed neutrinos 
the LHC bounds from $pp \to e+ MET + X$  are at the (sub)percent level,    
already stronger than those from $\beta$ decays. 
For operators  involving left-handed neutrinos, 
the (axial-)vector  and pseudo-scalar effective couplings are best probed at low energy, 
while scalar and tensor couplings  are currently probed at the same level by beta decays and the LHC channels $pp \to e+ MET + X$ and, by using SU(2) gauge invariance,  $pp \to e^+e^- + X$. 
Future beta decay experiments  at the 0.1\% level or better will compete in sensitivity with higher statistics and
higher  energy data from the LHC. 
\vfill

\end{titlepage}

\section{Introduction}

Beta decays  have  played a central role 
in determining the $V-A$ structure of charged current (CC)  weak interactions and in shaping 
what we now call the Standard Model (SM)~\cite{Weinberg:2009zz,Severijns2006dr}. 
Nowadays, precision beta-decay  measurements 
with neutrons, nuclei, and mesons~\cite{Severijns2006dr} 
can be used to probe the existence of non-standard  CC interactions 
that  would induce  effective violations 
of Cabibbo universality and lepton universality, and  distinctive 
non-V-A  signatures  in   the decay spectra and correlations. 

Low-energy  CC  processes  are sensitive to many classes of SM extensions 
and various mechanisms (tree level mediation by  novel vector or scalar bosons, 
vertex corrections,  box diagrams, etc)~\cite{Herczeg2001vk}. 
For example, for recent discussions of supersymmetric contributions see  Refs.~\cite{Profumo:2006yu,Bauman:2012fx}.
More generally, the new physics reach of low-energy beta decays can be studied in a 
model-independent way  within an effective field theory (EFT) setup, in which the dynamical 
effects of new heavy BSM  degrees of freedom are  parameterized by local operators, built with SM fields, and   
of dimension higher than four.
In this sense,   0.1\%   level (or better)   beta decay measurements  provide  ``broad band"  probe of  BSM interactions,
enforcing powerful low-energy  boundary conditions on virtually any SM extension at the TeV scale, 
currently probed  at the Large Hadron Collider (LHC). 

In this letter we  wish  to compare  the  new physics reach of low-energy 
beta decay measurements  and $pp$ collisions at the LHC in 
constraining non-standard CC interactions.  
Such a comparison is in principle  model-dependent, since it requires 
knowing the detailed dynamics of the SM extension. 
However,  if the particles that mediate the new interactions are above 
threshold for production at colliders, then the EFT analysis  
is valid at  collider energies  and a direct comparison of low-energy and 
collider constraints can be performed.  

Working in the latter scenario,  we present  in Section~\ref{sect:2}  the complete set of 
$SU(2) \times U(1)$-invariant operators that 
contribute to CC processes,   involving  the SM fields and  right-handed singlet neutrinos. 
We then discuss  the evolution of the effective Lagrangian down to hadronic scales  
and the matching to a nucleon-level effective theory. 
In Section~\ref{sect:low-energy}  we  summarize  the current bounds from low-energy 
and briefly discuss the reach of future beta decay experiments. 
In Section~\ref{sect:collider} we compute the bounds on 
non-standard interactions arising from the processes 
$pp \to e +MET + X$  and $pp \to e^+ e^- + X$ at the LHC, 
and show that the combination of low-energy and LHC searches 
provides a much stronger set of constraints  on non-standard  CC interactions. 
We present our final comments in Section~\ref{sect:conclusions}.

\section{EFT Analysis of Charged Current Processes}
\label{sect:2}

\subsection{Weak-scale Operator Basis}
The building blocks to construct gauge-invariant  local operators are the  gauge fields $G_\mu^A,  \, W_\mu^a, \, B_\mu$, corresponding to $SU(3)\times SU(2)_L \times U(1)_Y$, the six fermionic gauge multiplets (now including a singlet right-handed neutrino state),
\beq
q^i =
\left(
\begin{array}{c}
u_L^i \\
d_L^i
\end{array}
\right)  \qquad
u^i = u_R^i \qquad
d^i = d_R^i \qquad 
l^i =
\left(
\begin{array}{c}
\nu_L^i\\
e_L^i
\end{array}
\right)
\qquad e^i = e_R^i
\qquad \nu^i = \nu_R^i~,  
\label{eq:fermions}
\eeq
the Higgs doublet $\varphi$
\beq
\varphi =
\left(
\begin{array}{c}
\varphi^+ \\
\varphi^0
\end{array}
\right)~,
\eeq
and the covariant derivative
\beq
D_{\mu} =   I \, \partial_\mu   \, - \, i g_s \frac{\lambda^A}{2} G_\mu^A \,
- \, i g \frac{\sigma^a}{2} W_\mu^a   \, - \, i g'  Y B_\mu~.
\eeq
In the above expression  $\lambda^A$ are the $SU(3)$ Gell-Mann matrices, $\sigma^a$ are the $SU(2)$  Pauli matrices,  $g_s, g, g'$ are the gauge couplings and $Y$ is the hypercharge of a given multiplet.

The minimal and complete  set of dimension-six operators contributing to low-energy semi-leptonic charged current processes can be divided into two groups (operators involving the singlet  $\nu$ are displayed on the right columns below). \\
\noindent{\underline {Four-fermion operators}}:
\begin{subequations}
\bea
&& 
O_{l q}^{(3)}= (\overline{l} \gamma^\mu \sigma^a l) (\overline{q} \gamma_\mu \sigma^a q)~~~~~~~~~~~~~~~~~~~
O_{e\nu ud} = (\overline{e} \gamma^\mu \nu) (\overline{u} \gamma_\mu  d)
+ {\rm h.c.}~~~~~~~~
\\
&&
O_{qde} = (\overline{\ell} e) (\overline{d} q)+ {\rm h.c.}~~~~~~~~~~~~~~~~~~~~~~~
O_{qu\nu} = (\overline{\ell} \nu) (\overline{u} q)+ {\rm h.c.}~~~~~~~~~~
\\
&&
O_{l q} = (\bar{l}_a e)\epsilon^{ab}(\bar{q}_b u)+ {\rm h.c.}~~~~~~~~~~~~~~~~~~~~~
O_{l q} '  = (\bar{l}_a \nu)\epsilon^{ab}(\bar{q}_b d)+ {\rm h.c.}~~~~~
\\
&&
O^t_{l q} =(\bar{l}_a\sigma^{\mu\nu}e)\epsilon^{ab}(\bar{q}_b\sigma_{\mu\nu}u)+ {\rm h.c.}~~~~~~~~~~~~
O^{t'}_{l q} =(\bar{l}_a\sigma^{\mu\nu} \nu)\epsilon^{ab}(\bar{q}_b\sigma_{\mu\nu}d)+ {\rm h.c.}~~~~~~~~~~
\eea
\label{4fermi-operators}
\end{subequations}

\noindent{\underline{Vertex corrections}:
\begin{subequations}
\bea
&& O_{\varphi \varphi} = i(\varphi^T \epsilon  D_\mu \varphi) (\overline{u}\gamma^\mu d)+ {\rm h.c.}~~~~~~~~~ 
      O_{\varphi \varphi} ' = i(\varphi^T \epsilon  D_\mu \varphi) (\overline{\nu}\gamma^\mu e)+ {\rm h.c.}~~~~~~~~~
\\
&& 
O_{\varphi q}^{(3)} = \! i (\varphi^\dagger D^\mu \sigma^a \varphi)(\overline{q} \gamma_\mu \sigma^a q)+\!{\rm h.c.} ~~~~~~ 
\\
&&
O_{\varphi l}^{(3)} = \! i (\varphi^\dagger D^\mu \sigma^a \varphi)(\overline{l} \gamma_\mu \sigma^a l)+\!{\rm h.c.}~~~~~~
\eea
\label{vertexcorr-operators}
\end{subequations}


Denoting with  $\Lambda_i$  the effective dimensionful coupling associated with the operator $O_i$, we can write the effective Lagrangian as 
\bea
{\cal L}^{(\rm{eff})}
= {\cal L}_{\rm{SM}} + \sum_{i} \frac{1}{\Lambda_i^2}~ O_i \ \longrightarrow \ 
{\cal L}_{\rm{SM}} +  \frac{1}{v^2}  \, \sum_{i}  \, \hat{\alpha}_i   ~ O_i \, ,
\qquad 
{\rm with}  \  \  \hat{\alpha}_i = \frac{v^2}{\Lambda_i^2}~, 
\eea
where in the last step we have set the correct dimensions by the Higgs VEV $v = \langle \varphi^0 \rangle =   (2 \sqrt{2} G_F)^{-1/2}$ and defined the dimensionless  new-physics couplings $\hat{\alpha}_i$, which in general are matrices in both quark and lepton flavor spaces. 

\subsection{Low-scale Effective Lagrangian}

In this framework one can derive the low-scale  $O(1 \ {\rm GeV})$  effective Lagrangian for semi-leptonic transitions. It receives contributions from both $W$-exchange diagrams (with modified $W$-fermion couplings) and the four-fermion operators. This matching procedure leads to
\bea
{\cal L}_{\rm CC}  &=&
- \frac{G_F^{(0)} V_{ud}}{\sqrt{2}} \  \Big[ \ \Big( 1 +  \eL \Big) \  
\bar{e}  \gamma_\mu  (1 - \gamma_5)   \nu_{\ell}  \cdot \bar{u}   \gamma^\mu  (1 - \gamma_5)  d   
\label{eq:leff10}
 \\
& + &
\teL  \ \ \bar{e}  \gamma_\mu  (1 + \gamma_5)   \nu_{\ell}  \cdot \bar{u}   \gamma^\mu  (1 - \gamma_5)  d  
\nonumber\\
&+&   \eR   \  \   \bar{e}  \gamma_\mu  (1 - \gamma_5)   \nu_{\ell} 
\cdot \bar{u}   \gamma^\mu  (1 + \gamma_5)  d  
\  + \ 
\tilde{ \epsilon}_R   \  \   \bar{e}  \gamma_\mu  (1 +  \gamma_5)   \nu_{\ell} 
\cdot \bar{u}   \gamma^\mu  (1 + \gamma_5)  d  
\nonumber\\
&+&  \eS  \  \  \bar{e}  (1 - \gamma_5) \nu_{\ell}  \cdot  \bar{u} d 
 \ + \  \teS  \  \  \bar{e}  (1 +  \gamma_5) \nu_{\ell}  \cdot  \bar{u} d 
 \nonumber \\
 &-& \eP  \  \   \bar{e}  (1 - \gamma_5) \nu_{\ell}  \cdot  \bar{u} \gamma_5 d 
 \ -  \  \teP  \  \   \bar{e}  (1 + \gamma_5) \nu_{\ell}  \cdot  \bar{u} \gamma_5 d 
 \nonumber \\
 &+& 
\eT    \   \bar{e}   \sigma_{\mu \nu} (1 - \gamma_5) \nu_{\ell}    \cdot  \bar{u}   \sigma^{\mu \nu} (1 - \gamma_5) d
\ + \ 
\teT      \   \bar{e}   \sigma_{\mu \nu} (1 + \gamma_5) \nu_{\ell}    \cdot  \bar{u}  
 \sigma^{\mu \nu} (1 + \gamma_5) d~. 
\nonumber
\eea
Here $e,u,d$ denote the electron, up- and down-quark  mass eigenfields, while $\nu_\ell$ represents the neutrino flavor fields, with in general $\ell \neq e$. In what follows,  we suppress lepton flavor indices. Whenever observables involving neutrinos are considered, a summation over the (unobservable) neutrino flavors is assumed. The non-standard effective couplings $\epsilon_i$ and $\tilde{\epsilon}_i$ are given in terms of the weak-scale couplings $\hat{\alpha}_j$ as follows~\footnote{We split the correction $\epsilon_L$ to the SM operator into a vertex correction $\eLv$ and a contact correction $\eLc$ originating from a four-fermion weak scale operator. In the vertex correction $\eLv$ we include both linear and quadratic new physics effects.}: 
\begin{subequations}
\bea
\eL &=&  \eLv  +   \eLc 
~~~~~~~~~~~~~~~~~~~~~~~~~~~~~~~~~~~~~~~~~~~
\teL =  - \hat{\alpha}^{'*}_{\varphi \varphi}
 \\
\eLv &=& 2 \hat{\alpha}_{\varphi l}^{(3)} + 
 2 \frac{ [V \, (\hat{\alpha}_{\varphi q}^{(3)})^\dagger ]_{11}}{V_{ud} }
+ 4  \hat{\alpha}_{\varphi l}^{(3)} \, \frac{ [V \, (\hat{\alpha}_{\varphi q}^{(3)})^\dagger ]_{11}}{V_{ud} }
 \\
\eLc  &=&  - 2 \frac{ [V \, \hat{\alpha}_{lq}^{(3)} ]_{11}}{V_{ud} }
\\
\eR  &=&  -  \frac{ [\hat{\alpha}_{\varphi \varphi}]_{11}}{V_{ud} }
~~~~~~~~~~~~~~~~~~~~~~~~~~~~~~~~~~~~~~~~~~
\teR = - \frac{ \left[\hat{\alpha}_{e \nu u d}\right]_{11}}{V_{ud}} 
\\
\eS - \eP  &=&  - 2 \frac{ [V \, \hat{\alpha}_{qde}^\dagger ]_{11}}{V_{ud} }
~~~~~~~~~~~~~~~~~~~~~~~~~~~~~~~
\teS - \teP  =   2 \frac{ [V \, \hat{\alpha}_{l q} ' ]_{11}}{V_{ud} }
\\
\eS + \eP  &=&  - 2 \frac{ [\hat{\alpha}_{l q}^\dagger ]_{11}}{V_{ud} }
~~~~~~~~~~~~~~~~~~~~~~~~~~~~~~~~~~~
\teS + \teP  =   - 2 \frac{ [ \hat{\alpha}_{qu \nu} ]_{11}}{V_{ud} }
\\
\eT  &=&  -  \frac{ [\hat{\alpha}_{l q}^{t \ \dagger}]_{11}}{V_{ud} }
~~~~~~~~~~~~~~~~~~~~~~~~~~~~~~~~~~~~~~~~~~
\teT  =    \frac{ [V \, \hat{\alpha}_{l q}^{t \ '} ]_{11}}{V_{ud} }~.
\eea
\label{eq:match1}
\end{subequations}
In the above matching conditions $V$ denotes the CKM matrix, and the quark family indices ``$11$" are explicitly displayed. The $\alpha$ coefficients are defined in a flavor basis where both the down-quark and charged lepton Yukawa matrices are diagonal.

We note here that while the physical amplitudes  are  renormalization scale  and scheme independent, the individual effective couplings 
 $\epsilon_{S,P,T}$ ($\tilde{\epsilon}_{S,P,T}$) and the corresponding  hadronic matrix elements can display a strong scale dependence. Throughout the paper, we will quote estimates and bounds for the $\epsilon_i$ ($\tilde{\epsilon}_i$) at the renormalization scale  $\mu=2$~GeV in the $\overline{\rm MS}$ scheme.

\subsection{Nucleon-level effective couplings}

The quark-level effective Lagrangian  (\ref{eq:leff10}) can be matched onto a nucleon-level 
effective Lagrangian by computing the 
one-nucleon matrix elements of all quark  bilinears. 
To leading order in momentum transfer one 
has $\langle p | \bar{u} \Gamma d | n \rangle = g_\Gamma  \, \bar{\psi}_p \Gamma \psi_n$  with $\Gamma = 1, \gamma_5, \gamma_\mu, \gamma_\mu \gamma_5, \sigma_{\mu \nu}$, 
and  one  then recovers the  Lee-Yang~\cite{Lee:1956qn}  effective Lagrangian. 
The Lee-Yang~\cite{Lee:1956qn}  effective couplings $C_i$, $C_i'$ ($i \in \{V,A,S,T\}$) can be expressed in terms of our parameters as 
\begin{subequations}
\label{eq:matchLY}
\bea
C_{i} &=& \frac{G_F}{\sqrt{2}} \,  V_{ud} \,  \bar{C}_{i}    \\
\bar{C}_V &=& g_V  \left(1 + \eL + \eR  
+ \teL + \teR 
\right)  \\
\bar{C}_V' &=& g_V  \left(1 + \eL + \eR  
- \teL - \teR 
\right)  \\
\bar{C}_A &=& - g_A  \left(1 + \eL - \eR 
- \teL   +  \teR  
\right)  \\
\bar{C}_A' &=& - g_A  \left(1 + \eL - \eR 
+ \teL   -  \teR  
\right)  \\
\bar{C}_S &=&   g_S  \, \left(  \eS  + \teS \right) \\
\bar{C}_S'  &=&   g_S  \, \left(  \eS  -  \teS \right) \\
\bar{C}_P &=&   g_P  \, \left(  \eP  - \teP \right) \\
\bar{C}_P'  &=&   g_P  \, \left(  \eP  +  \teP  \right) \\
\bar{C}_T &=&    4 \, g_T \, \left( \eT   + \teT \right)  \\
\bar{C}_T '  &=&    4 \, g_T \, \left( \eT   -  \teT \right) ~.  
\eea
\end{subequations}
Using these relations and the results of Ref.~\cite{Jackson1957zz} one can work out the dependence of
neutron and nuclear  beta decay observables on the short-distance parameters $\epsilon_i$ and $\tilde{\epsilon}_i$. 

The vector charge $g_V=1$ up to tiny second-order isospin-breaking corrections. 
First principle calculations of $g_{A,S,P,T}$ are  nowadays possible with lattice QCD (LQCD). 
The status of LQCD calculations of these charges is critically reviewed in~\cite{Bhattacharya:2011qm}, 
where the first estimate of $g_S$  from LQCD was provided.
For the axial coupling $g_A$, 
the upshot is that   different calculations give results in the range $1.12 < g_A < 1.26$, 
with errors much larger than the one achieved in the experimental determination~\cite{Nakamura:2010zzi}.
New lattice analyses of the  scalar  and tensor charge have been recently performed. 
Ref.~\cite{Bhattacharya:2011qm} reported  $g_S = 0.8 \pm 0.4$ and $g_T = 1.05 \pm 0.35$
in the $\overline{MS}$ scheme and at $\mu=2$~GeV,  
where the uncertainties include an estimate of all the systematic effects (volume, lattice spacing, and chiral extrapolations). 
More recently Ref.~\cite{Green:2012ej} provided improved results for the scalar 
and tensor charges,  $g_S = 1.08 \pm 0.28$ and $g_T = 1.038 \pm 0.011$, 
with uncertainty associated with statistics and chiral extrapolation. 
These results, however,   do not include an estimate of the 
systematic error associated with finite volume and finite lattice spacing extrapolations. 
Therefore, in what follows we will still use  the results of Ref.~\cite{Bhattacharya:2011qm} 
as baseline lattice results.

\section{Low energy bounds}
\label{sect:low-energy}

The couplings $\epsilon_\alpha$ and $\tilde{\epsilon}_\beta$ 
appearing in the Lagrangian (\ref{eq:leff10}) 
have been traditionally constrained through 
low-energy probes, such as decays of the pion, the neutron, and nuclei. 
Detailed analyses of these low-energy bounds can be found in 
Refs.~\cite{Herczeg2001vk,
Severijns2006dr,Cirigliano:2009wk,Bhattacharya:2011qm}.
Here we  present a brief  survey of current and future constraints  using the notation introduced in the previous section. 

The ten non-standard couplings can be divided in two classes: 
the  $\epsilon_\alpha$ that involve L-handed neutrinos,  and the $\tilde{\epsilon}_\beta$  that involve R-handed neutrinos.  The $\tilde{\epsilon}_\beta$ appear in decay rates and distributions  either quadratically or multiplied by the small factor $m_\nu/E_\nu$ (through interference of the SM and BSM couplings). On the other hand, the $\epsilon_\alpha$ couplings contribute to decay rates to linear order without any  suppression. As a consequence, the low-energy bounds on the $\epsilon$'s will be stronger than the bounds on the $\tilde{\epsilon}$'s. 

Experiments constrain the products of hadronic matrix elements and quark-level NP couplings $\epsilon_i$ and $\tilde{\epsilon}_i$, or linear combinations of these products. The confidence intervals on the $\epsilon_i$ and $\tilde{\epsilon}_i$  are obtained using the so-called R-Fit method~\cite{Charles:2004jd}, which assumes that the matrix elements are bound to remain within allowed ranges determined by the analytical or numerical calculations.\footnote{In the case at hand the relevant ranges are $0.4 \leq  g_S  \leq 1.2$ and  $0.7 \leq g_T \leq 1.4$ for the nucleon matrix elements \cite{Bhattacharya:2011qm} and $0.20\leq f_T \leq 0.28$ for meson form factor appearing in the radiative pion decay \cite{Mateu:2007tr}.}
We now discuss in turn the low-energy  bounds on the $\epsilon$'s  and $\tilde{\epsilon}$'s,  and summarize the results  in Tables~\ref{tab:summaryL}  and \ref{tab:summaryR}. 
\\

\noindent{\bf Vector and axial couplings}: 
The combinations $(\epsilon_L \pm \epsilon_R)$ affect  the overall normalization of the effective Fermi constant  in processes mediated by the vector and axial-vector current, respectively. 
The hadronic matrix elements of the vector current are known very precisely through QCD  flavor symmetry considerations (SU(2) and SU(3)), while the axial-vector matrix elements require  non-perturbative calculations. 
Therefore,  while the combination $(\epsilon_L-\epsilon_R)$ remains relatively poorly unconstrained, $(\epsilon_L+\epsilon_R)$ is strongly constrained by quark-lepton universality tests (or CKM unitarity tests)~\cite{Cirigliano:2009wk}, that involve  a precise determination of $V_{ud}$  and $V_{us}$ from processes mediated by the vector current (such as  $0^+ \to 0^+$ nuclear decays and $K \to \pi \ell \nu$).
Assuming that one operator at the time dominates, CKM unitarity constraints impose severe bounds on $\epsilon_L^{(v)}$,   $\epsilon_L^{(c)}$, $\epsilon_R$. The bound on the vertex correction $\epsilon_L^{(v)}$ is  comparable to the one  obtained by $Z$-pole observables~\cite{Cirigliano:2009wk}, while the bounds on   $\epsilon_L^{(c)}$ and  $\epsilon_R$ are stronger than from any  other source. 
Independent handles on $\epsilon_R $ are not possible because neutron and nuclear correlation decay experiments are only sensitive to the combination $(1 - 2 \epsilon_R)g_A/g_V$, 
so that  disentangling $\epsilon_R$ requires precision measurements of $(1 - 2 \epsilon_R)g_A/g_V$ and precision calculations of $g_A/g_V$ in LQCD, which  are not yet at the required sub-percent level. 
\\

\noindent{\bf Pseudoscalar coupling}: 
The effective pseudoscalar coupling $\epsilon_P$ contributes to leptonic decays of the pion $\pi \to \ell \nu_\ell$. Strong constraints are implied by the helicity-suppressed ratio $R_\pi \equiv \Gamma(\pi \to e \nu [\gamma])/\Gamma(\pi \to \mu \nu [\gamma])$~\cite{Britton:1992pg,Czapek:1993kc,Cirigliano:2007xi},  assuming that the coupling  $\epsilon_P$ is independent of the lepton masses (in particular that they do not satisfy  $\epsilon_P^{(e)}/m_e = \epsilon_P^{(\mu)}/m_\mu$). Even after taking into account possible cancellations between flavor diagonal and non-diagonal contributions one obtains the strong bound reported in Table~\ref{tab:summaryL} (for details see \cite{Bhattacharya:2011qm}).
\\

\noindent{\bf Scalar and tensor  couplings}: 
The scalar  and tensor couplings  $\epsilon_S$ and  $\epsilon_T $ contribute to  linear order  to the Fierz interference terms in beta decays of neutrons and nuclei, and to  the neutrino-asymmetry correlation coefficient $B$ in polarized neutron and nuclear decays. Because of the peculiar way in which the Fierz interference term appears in many asymmetry measurements, bounds on $\epsilon_S$ and $\epsilon_T$ can also be obtained by observation of the beta-asymmetry correlation coefficient $A$,  electron-neutrino correlation $a$,   and positron polarization measurements in various nuclear beta decays.  
Currently, the most stringent constraint on $\epsilon_S$ arises from the Fierz interference term in  $0^+ \to 0^+$  nuclear beta decays ($   -0.001  <  g_S \epsilon_S <  0.0032$ at 90\% CL)~\cite{Hardy:2008gy}, while the strongest constraint on $\epsilon_T$  arises from a Dalitz-plot analysis of the radiative pion decay $\pi \to e \nu \gamma$~\cite{Bychkov:2008ws,Mateu:2007tr} ($ - 2.0 \times 10^{-4}  \ < \ \epsilon_T \,  f_T \  <  \   2.6  \times 10^{-4} $ at 90\% CL).

Moreover, as discussed  in Refs.~\cite{Voloshin:1992sn,Herczeg:1994ur, Campbell:2003ir} and summarized in \cite{Bhattacharya:2011qm}, there are also potentially very strong constraints on $\epsilon_{S,T}$  from $R_\pi \equiv \Gamma(\pi \to e \nu [\gamma])/\Gamma(\pi \to \mu \nu [\gamma])$, due to operator mixing: once a scalar or tensor interaction is generated by new physics, SM radiative corrections will generate an  effective pseudoscalar operator that mediates the helicity-suppressed modes. Numerically, the bounds are at the level of $|\epsilon_S| \lesssim 8  \times 10^{-2}$ and $|\epsilon_T| \lesssim  10^{-3}$, almost as strong as the 	``direct" bounds discussed above. It is important to keep in mind, however, that if the  flavor structure of the SM extension is known, this constraint could be the strongest.

In the future, $10^{-3}$-level measurements of the antineutrino asymmetry in neutron decay~\cite{WilburnUCNB,abBA}, and the Fierz interference term in neutron decay~\cite{Pocanic:2008pu,UCNb} and  in the pure Gamow-Teller  decay of $^6$He~\cite{Knecht201143} will probe $\epsilon_T$ in the  $5 \times 10^{-4}$ range. 
\\

\noindent{\bf  
Couplings involving R-handed neutrinos ($\tilde{\epsilon}_\alpha$)}: 
Neglecting neutrino masses, all the $\tilde{\epsilon}_\beta$ couplings contribute to decay rates incoherently with the SM, i.e. $\propto | \tilde{\epsilon}_\beta|^2$. Detailed expressions of the contributions to neutron and nuclear beta decay correlation coefficients can be found  in~\cite{Jackson1957206} (one needs to re-express the Lee-Yang couplings in terms of the  $\epsilon_\alpha$ and  $\tilde{\epsilon}_\beta$ using  Eqs~\ref{eq:matchLY}  above).
The corresponding bounds can be obtained  from the analysis of  Ref.~\cite{Severijns2006dr}, in particular from the fits to beta decay data allowing for non-zero $\teL, \teR$ only,  
implying $|\teL \pm \teR| < 0.06$, and  $\teS, \teT$ only, implying $|g_S \teS| < 0.06$,  $|g_T \teT| <  0.02$ at 90\% CL.

The ratio of leptonic pion decay rates $R_\pi \equiv \Gamma ( \pi \to e \nu [\gamma]) / \Gamma ( \pi \to \mu \nu [\gamma])$ 
is a very sensitive probe of the pseudoscalar coupling $\teP$ 
and through operator mixing   can also constrain $\teS$, $\teT$. 
Assuming possible cancellations between $\epsilon_P$ and $\teP$ 
(see discussion in Ref.~\cite{Bhattacharya:2011qm} with the replacement $\epsilon_P^{ex} \to \teP^{ee}$),  
one obtains $|\teP| < 2.8 \times 10^{-4}$. 
The  mixing  constraint can be worked out by using the three-operator mixing results from Ref.~\cite{Campbell:2003ir}, namely:
\footnote{There is a typo in Eqs. (25-26) of  Ref.~\cite{Campbell:2003ir},  
where the definitions of the operators $O_1$ and $O_2$  should be interchanged.}
\begin{subequations} 
\bea
\teP (\mu) &=& \teP  (\Lambda) \left(1 + \gamma_{PP} \,  \log \frac{\Lambda}{\mu} \right) 
+ \teS  (\Lambda)  \ \gamma_{SP} \,  \log \frac{\Lambda}{\mu} 
+ \teT  (\Lambda)\  \gamma_{TP} \,  \log \frac{\Lambda}{\mu} 
\\
\gamma_{PP} &=& \frac{3}{4} \frac{\alpha_2}{\pi}  + \frac{49}{144} \frac{\alpha_1}{\pi}  \approx 8.6 \times 10^{-3}
\\
\gamma_{SP} &=&  \frac{3}{16} \frac{\alpha_1}{\pi}  \approx 0.6 \times 10^{-3}
\\
\gamma_{TP} &=& \frac{9}{2} \frac{\alpha_2}{\pi}  -  \frac{1}{3} \frac{\alpha_1}{\pi} \approx +4.4 \times 10^{-2}~,
\eea
\end{subequations}
where $\alpha_1 = \alpha/\cos^2\theta_W$ and $\alpha_2 = \alpha/\sin^2\theta_W$ are the $U(1)$ and $SU(2)$ weak couplings, expressed in terms of the fine-structure constant and the weak mixing angle.  Setting $\teP (\Lambda) = 0$  and neglecting the small $O(\alpha/\pi)$ fractional difference between $\tilde{\epsilon}_{S,T} (\Lambda)$ and the observable $\tilde{\epsilon}_{S,T} (\mu)$ at the low scale, the 90\% C.L. constraint on the $\teS$-$\teT$ plane reads 
\beq
| \, 0.6  \ \teS  \ + \  44.2 \ \teT  |  \le  \frac{0.28}{\log (\Lambda/\mu)}~.
\eeq
Assuming $\log (\Lambda/\mu) \sim 10$ (e.g. $\Lambda \sim 10\, {\rm TeV}$ and $\mu \sim  1 \, {\rm GeV}$),  we get $| \teS |  \lesssim  5 \times 10^{-2}$ and $| \teT |  \lesssim 0.6 \times 10^{-3}$, which are the strongest  low-energy  bounds on the $\tilde{\epsilon}$'s couplings. However, given the dependence of these results  on assumptions about the flavor structure of the couplings, we do not report these bounds in Table~\ref{tab:summaryR}. It is worth pointing out that the LHC bound on $\teS$ derived in the next section and shown in Table~\ref{tab:summaryR} is a factor four stronger than this bound from $R_{\pi}$.

\section{Collider bounds}
\label{sect:collider}

Collider searches can also  probe the various non-standard couplings $\epsilon_\alpha$  and $\tilde{\epsilon}_\beta$ defined above. In Ref.~\cite{Bhattacharya:2011qm}   bounds were derived on $\epsilon_{S,T}$ by analyzing LHC data in the $pp  \to  e +MET+ X$ channel at $\sqrt{s} =7$~TeV and 1 fb$^{-1}$ integrated luminosity. 
Here we extend the analysis to all non-standard charged-current couplings and we consider bounds from both $pp  \to  e +MET + X$ and $pp  \to e^+ e^-  + X$  channels (using SU(2) gauge invariance). 

As in Ref.~\cite{Bhattacharya:2011qm}, here our analysis is to lowest order (LO) in the QCD interactions. While QCD effects have a small impact on the transverse mass distribution \cite{Barger:1983wf}, their impact on the overall normalization ($K$-factor) and dilepton invariant mass distributions of the NP interactions are not known and affects our analysis at the order $\alpha_s/\pi$. To maintain an analysis consistent to the same order in the QCD interactions, we therefore do not include a $K$-factor in our projections of the background. For our analyses of existing data we use background estimates provided by the experimental collaborations.

\subsection{The   $pp  \to  e +MET + X$ channel}

This channel is directly related to beta decays, since the parton-level process is $\bar{u} d \to  e \bar{\nu}$.  
We will use the (cumulative) transverse mass distribution to put bounds on the non-standard interactions. 
Including the $W$-exchange Standard Model contribution, to leading order (LO) in QCD the  $ p p \to e \nu + X$ 
cross-section for lepton transverse mass  $m_{T} \equiv \sqrt{2 E^e_T E^\nu_T (1 - \cos \Delta \phi_{e\nu})}$
greater than a threshold $\overline{m}_T$ takes the following form\footnote{The interference of (pseudo)scalar and tensor interactions vanishes after integration over the final leptons rapidities if the integration region is symmetric (under the exchange of both rapidities), as it happens in the absence of any cut. However experimental cuts break this symmetry (see later in this section) generating a residual contribution from these interference terms. We find the relative contribution to be numerically small.}:
\bea
\sigma(m_T > \overline{m}_T) &=&    
\sigma_W \  \Big[ ( 1 +   \eLv)^2+  |\teL|^2  + | \eR|^2 \Big]
- 2 \, \sigma_{WL}  \ \eLc \left( 1 +  \eLv \right) \nonumber\\
&+& \sigma_R  \ \Big[ |\teR|^2 + |\eLc|^2 \Big]
+ \sigma_S \  \Big[ |\eS|^2  +  |\teS|^2  + |\eP|^2 +  |\teP|^2   \Big]\nonumber\\
&+& \sigma_T  \  \Big[ |\eT|^2  +  |\teT|^2     \Big]~, 
\label{eq:sigmamt}
\eea 
where,
defining $\bar{\tau} = \overline{m}_T^2 / s$, the individual contributions 
$ \sigma_i$ ($i \in \{W, WL, R, S,T\}$) read
\begin{subequations}
\bea 
\sigma_i &=&  \frac{|V_{ud}|^2}{192\pi} \frac{s}{v^4}   \ \int_{\bar{\tau}}^1 \ d \tau \, L(\tau) 
\sqrt{\tau(\tau - \bar{\tau})} \  g_i (\tau) \ \ 
\\
g_W (\tau) &=&  \frac{8}{3} \frac{M_W^4}{s^2}  \frac{1 - \frac{\bar{\tau}}{4 \tau}}{\left(\tau - \frac{M_W^2}{s}\right)^2  + \frac{\Gamma_W^2}{s}  \frac{M_W^2}{s}}   
\\
g_{WL} (\tau) &=& g_{W} (\tau) \ \frac{s}{M_W^2} \left(\tau - \frac{M_W^2}{s}\right)
\\
g_{R} (\tau) &=& \frac{8}{3}   \left( 1 - \frac{\bar{\tau}}{4 \tau} \right)
\\
g_{S} (\tau) &=& 1 
\\
g_{T} (\tau) &=&  \frac{32}{3}   \left( 1 - \frac{\bar{\tau}}{\tau} \right)
\\
L (\tau) &= & \int ^{-\frac{1}{2} \ln \tau}_{\frac{1}{2} \ln \tau} d y_P  \Big[ f_{\overline{u}}(\sqrt{\tau} e^{y_P}) f_d (\sqrt{\tau} e^{-y_P}) + (\overline{u},d) \rightarrow (u,\overline{d}) \Big] ~.
 \eea
\end{subequations}

\begin{center}
\begin{figure}[floatfix] 
\centering
\vspace{-2.5cm}
\includegraphics[width=0.80\textwidth]{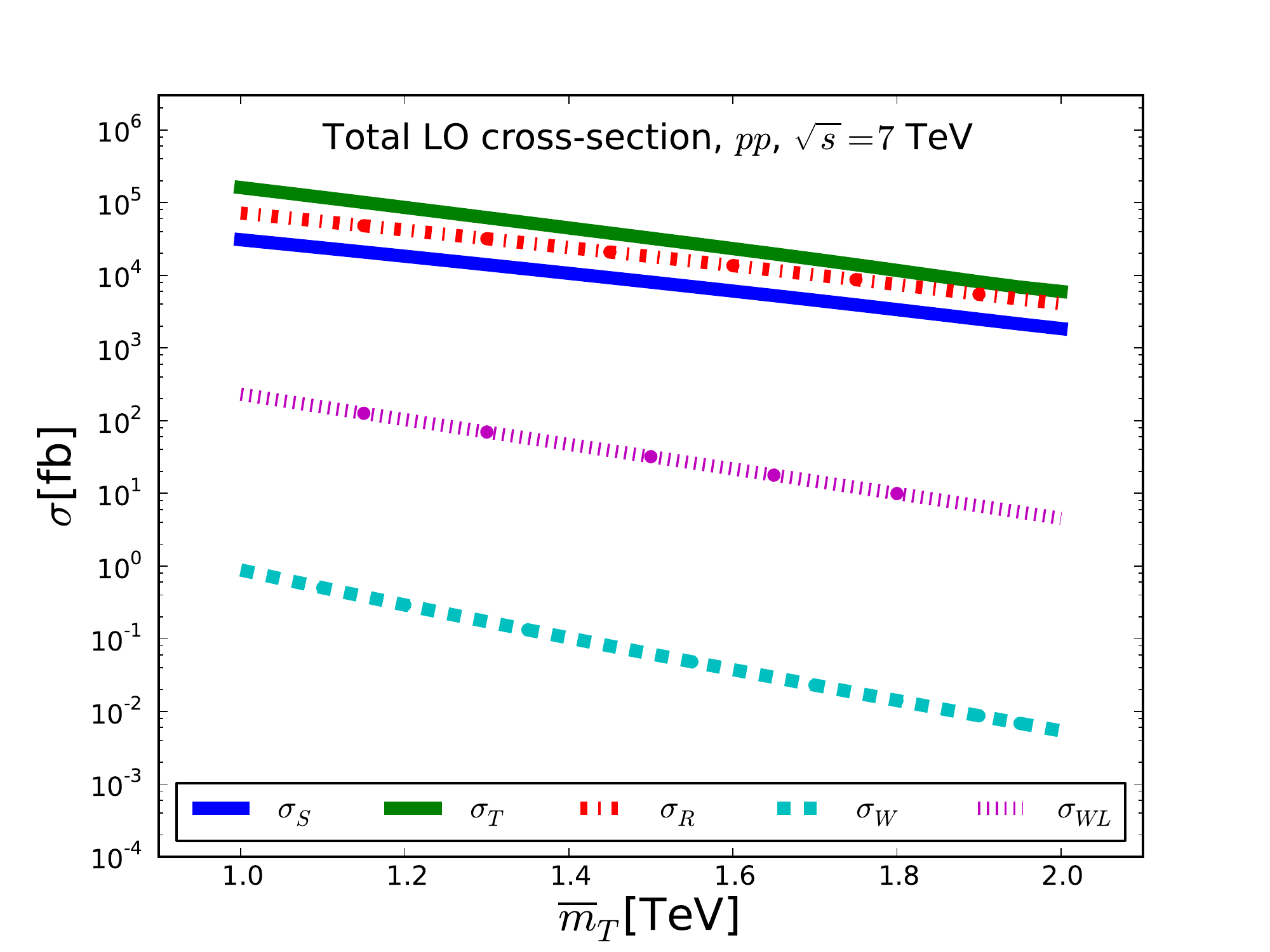}
\caption{\label{fig:sig1}   Cross-sections introduced in Eq.~(\ref{eq:sigmamt}) as a function of the cut on the transverse mass $\overline{m}_T$, for $\sqrt{s}=7$ TeV. We see that in the entire plot range we have the hierarchy $\sigma_T>\sigma_R>\sigma_S\gg\sigma_{WL}\gg\sigma_W$.
}
\end{figure}
\end{center}
The  various contributions  to $\sigma (m_T > \overline{m}_T)$ 
are plotted in Fig.~\ref{fig:sig1} using the CTEQ6 set of parton distribution functions (PDF) in the $\overline{MS}$ scheme, referred to as CTEQ6M, evaluated at $Q^2=1$ TeV$^2$ \cite{Pumplin:2002vw}. In the Figure we see a clear hierarchy that determines the sensitivity to the different kind of new interactions: the NP couplings that appear multiplying $\sigma_W$ in Eq.~\eqref{eq:sigmamt} will be essentially unconstrained by this analysis, since they do not have any enhancement with respect to the SM background, whereas for the rest of the NP couplings we obtain strong bounds. 
One important feature emerging from  Fig.~\ref{fig:sig1} is that the SM term $\sigma_W$ goes to zero faster than the rest of terms, 
thus  implying  that  larger values of the transverse mass cut  $\overline{m}_T$  increase the signal/background ratio. 

Writing $\sigma (m_T > \overline{m}_T) = \sigma_W  + \sigma_{NP} (\epsilon_\alpha)$, and using the current CMS and ATLAS results on  $pp  \to  e \nu + X$, one can put bounds on the couplings $\epsilon_\alpha$. For a given number of expected background events $n_b$ and actually observed events $n$, the $90 \%$ C.L. upper bound on the number of NP signal events $n_s^{up}$ can be calculated using \cite{Nakamura:2010zzi} 
\beq 
0.10 = e^{-n^{\rm{up}}_{\rm{s}}}  \cdot \frac{\sum_{m=0} ^{n} \frac{1}{m!}(n^{\rm{up}}_{\rm{s}}+n_{\rm{b}})^m}{\sum_{m=0} ^{n} \frac{1}{m!}n_{\rm{b}}^m}~,
\label{cl2}
\eeq 
which implies $\sigma_{NP}  <  n_s^{up} / (\epsilon_{eff}  \ L)$, where $\epsilon_{eff}$ represents the acceptance ($\epsilon_{A}$)~\footnote{
For the processes that we are studying, CMS performs the following cuts on the rapidities $y,y'$ of the charged leptons:
(i) $|y,y'| < 2.5$;  (ii) $|y,y'|\notin (1.442,1.560)$ (the transition region between barrel and endcaps);
and for the dilepton search, in addition (iii) CMS rejects events in which both leptons are in the endcap regions. \label{CMSacceptance}} 
times the lepton detection efficiency factor ($\epsilon_{det}$), and $L$ is the integrated luminosity.

We use for our study the most recent analysis of this channel published by the CMS Collaboration \cite{ENcms5fb}, with 5 fb$^{-1}$ at $\sqrt{s}=$ 7 TeV. The observed and expected number of events above the chosen cut on the transverse mass, and the associated upper bound on NP signal events $n_s^{up}$, are given in Table~\ref{tab:dataLHCenu}. Also shown are analogous numbers based on {20} fb$^{-1}$ of integrated luminosity at $\sqrt{s}=$ 8 TeV. To make this projection we assume no events are detected. This is a reasonable assumption, since  from a LO computation of $pp \rightarrow W^* \rightarrow e \nu$ we expect $\sim $ 0.3 background events for 
$\overline{m}_T = 2$~TeV.\\

\begin{table}[tbh]\centering
\begin{tabular}{|c c c c c c c c| }
\hline
					& Reference					&	$\sqrt{s}$ (TeV)	&	${\cal L}$ (fb$^{-1}$)	&	$\overline{m}_T$ (TeV)	&	$n$		&	$n_b$			&	$n_s^{up}$		\\
\hline
LHC-7  & CMS \cite{ENcms5fb}		&	7 				&	5.0					&	1.2					&	3		&	$2.8\pm 1.0$		&	4.5			\\	
LHC-8       & Projection					&	8				&	20					&	2.0					&	0		& 	$<1$			&	2.3			\\	
\hline
\end{tabular}
\caption{Experimental data (real or expected) and cuts over the transverse mass used in our analysis of $pp \to e +MET + X$. $n$  ($n_b$) is the number of observed  (background) events, and $n_s^{up}$ is the 90\% CL upper limit on the number of signal events. For the projection no events are assumed to be observed.
}
\label{tab:dataLHCenu}
\end{table}

The acceptance $\epsilon_{A}$ of the signal is set by the rapidity coverage of the detector and the kinematic cuts applied at the analysis-level. While the CMS analysis does have such additional cuts, to LO in the QCD interactions these have 100\% acceptance on the signal, such that the acceptance is almost entirely given by the geometric acceptance $\epsilon_{geom}$ of the lepton. 
We find a very high geometric acceptance, in the range 97-100\% (the lowest value being associated with the tensor-induced cross-section $\sigma_T$). The variance between interactions is at most a couple of percent, and the dependence on $\overline{m}_T$ or $\sqrt{s}$ is mild. The reason for these features is simply because for such high values of $\overline{m}_T$ the lepton is almost always central. 

Next, we infer from \cite{ENcms5fb} a lepton detection efficiency of $\sim 90\%$. Taken together, for our analysis we therefore use $\epsilon_{eff} \approx 0.87$. It is worth noting that varying $\epsilon_{eff}$ through the range of $0.7-1.0$, the NP bounds presented below are modified by only O(20\%), which is at the same level as the PDF and NLO QCD uncertainties. 

\begin{table}[tbh]\centering
\begin{tabular}{|c c c c c|}
\hline
				&	$|\epsilon_{S,P}|, |\tilde{\epsilon}_{S,P}|~~$	&		$|\eT|,  |\teT|$				&	$|\teR|$				&	$\eLc$						\\
\hline
LHC-7 			&	$1.3\times 10^{-2}$							&		$2.9\times 10^{-3}$			&	$4.9 \times 10^{-3}$	&	$(-3.1,+8.0)\times 10^{-3}$		\\
LHC-8  			&	$0.9\times 10^{-2}$							&		$2.2\times 10^{-3}$			&	$3.4\times 10^{-3}$	& 	$(-2.5,+4.6)  \times 10^{-3}$		\\
\hline
\end{tabular}
\caption{LHC bounds (first row) and projected bounds (second row) at 90\% C.L., on $\epsilon_i$ and $\tilde{\epsilon}_i$ ($i=S,P,T,R$) at the renormalization scale $\mu = 2$~GeV in the $\overline{\rm MS}$ scheme using the information given in Table \ref{tab:dataLHCenu}.}
\label{tab:resultsLHCenu}
\end{table}

Putting  the above results together, the bounds on the different NP coefficients are given in Table \ref{tab:resultsLHCenu}. These results are obtained using the CTEQ6M parton distribution functions \cite{Pumplin:2002vw} at $Q^2=1$ TeV$^2$. To estimate the uncertainty in these bounds arising from the parton distribution functions, we also used the MSTW2008 PDF sets \cite{Martin:2009iq}  and obtained bounds that differ by O(10\%) from those presented in the Table \ref{tab:resultsLHCenu}.
We also find comparable bounds from the the ATLAS $4.7$ fb$^{-1}$ $e+MET+X$ analysis at $\sqrt{s}=7$ TeV \cite{ATLAS:2012dm}, as well as from the CMS 3.7 fb$^{-1}$ analysis at $\sqrt{s}=8$ TeV \cite{CMS8TeV}.

Note that since collider searches set limits on the effective couplings $\epsilon_i$ and $\tilde{\epsilon}_i$ at the high renormalization scale $\mu = 1$~TeV,  a direct comparison with the low-energy constraints requires an appropriate rescaling down to the hadronic scale. Using the one-loop anomalous dimensions for the different quark bilinears (see~\cite{Broadhurst:1994se} and references therein), the one-loop beta function for the strong coupling constant, and including the appropriate heavy quark thresholds, we find in the $\overline{\rm MS}$ scheme $\epsilon_{S,P} (1~{\rm TeV})/\epsilon_{S,P} (2~{\rm GeV}) = 0.56$ and $\eT (1~{\rm TeV})/\eT (2~{\rm GeV}) = 1.21$, whereas $\epsilon_{L,R}$ are the same at both scales since the associated bilinears have zero anomalous dimension\footnote{The  $\tilde{\epsilon}_i$ couplings run under QCD in the same way as the $\epsilon_i$.}. These factors have already been taken into account in the results shown in Table \ref{tab:resultsLHCenu}.

These bounds have been obtained assuming contributions from 
one operator at a time and one might wonder about possible cancellations when several NP terms are present. For that matter, first of all one should notice that all the contributions to $\sigma_{NP}$ are positive-definite, except for those generated by $\epsilon_L^{(v,c)}$, where a negative contribution due to destructive interference with the SM amplitude (or among the NP terms) is possible. 
However, the constraints on these NP terms from LEP and CKM-unitarity tests are very strong: $|\epsilon_L^{(v,c)}| < 5\cdot 10^{-4}$ at 90\% C.L.~\cite{Cirigliano:2009wk}, and this leaves room only for a very small negative contribution in $\sigma_{NP}$, that would weaken our bounds on the other NP couplings at most by a few percent.

\begin{center}
\begin{figure}[!t] 
\centering
\vspace{-2.5cm}
\includegraphics[width=0.8\textwidth]{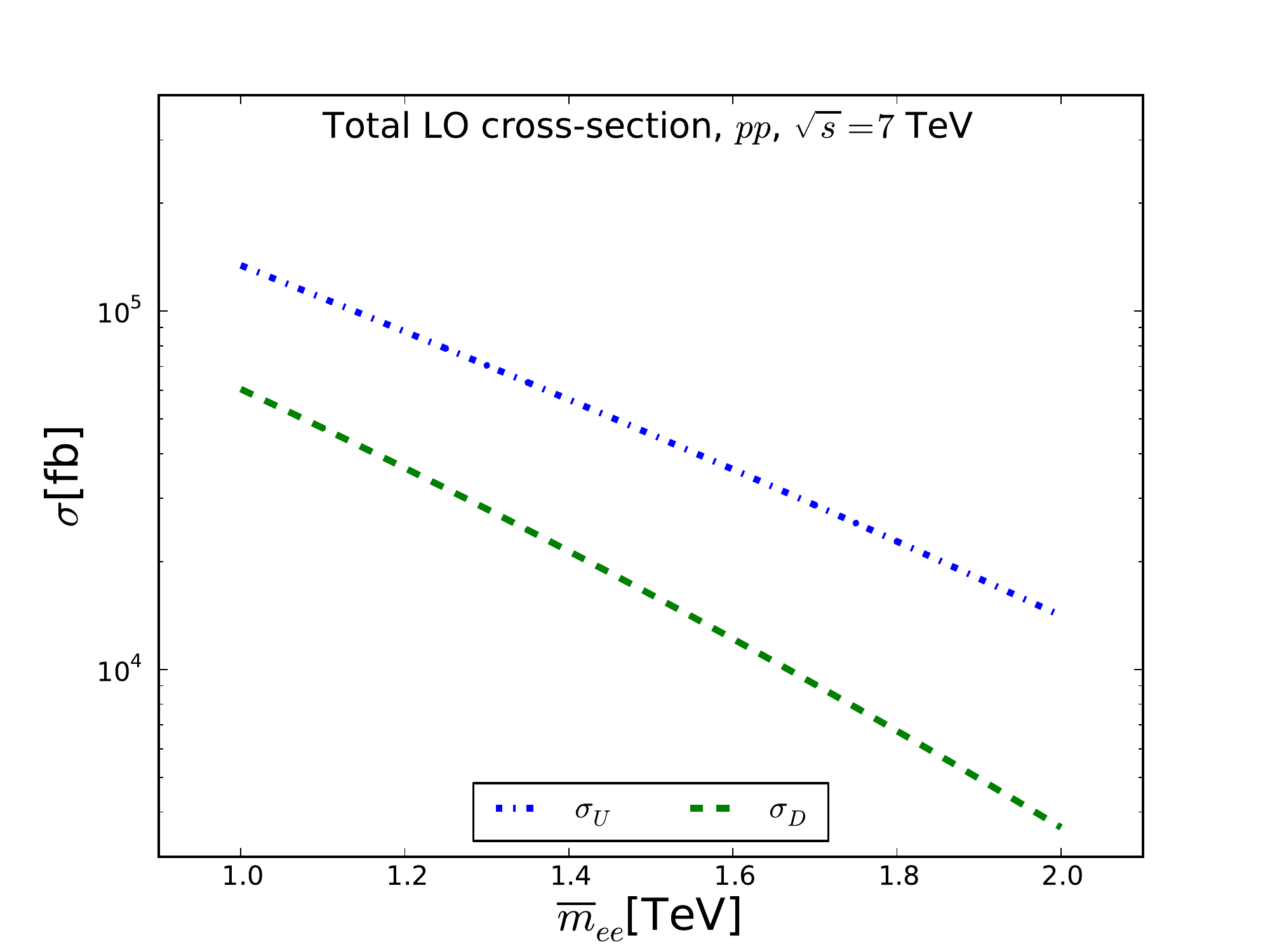}
\caption{\label{fig:sig2}   Cross-sections  $\sigma_{u,d}$ defined in Eq.~(\ref{eq:sigNC}) versus the cut $\overline{m}_{ee}$ over the dilepton invariant mass.
}
\end{figure}
\end{center}

\subsection{The $pp \to e^+ e^- + X$ channel}

$SU(2)$ gauge invariance implies that some of the 4-fermi operators (\ref{4fermi-operators}) and vertex correction operators (\ref{vertexcorr-operators}) mediate the transitions $q \bar{q} \to e^+ e^-$ and/or $q \bar{q} \to \nu \bar{\nu}$ ($q=u,d$). At the LHC the first transition contributes to the Drell-Yan process $pp  \to  e \bar{e} + X$, while the second one contributes to a mono-jet signal arising from the  parton-level process $q \bar{q} \to \nu \bar{\nu} + g$. 
Because this parton-level process involves one additional perturbative vertex, we expect weaker bounds from the mono-jets searches, and focus therefore on the Drell-Yan process  $pp \to e \bar{e} + X$. 

In what follows we additionally focus on those operators that involve a chirality-flip. Then the relevant interaction Lagrangian is given by
\beq
{\cal L}_{\rm NC}  =
\frac{\delta_{P-S}}{v^2} \ \bar{e}_R e_L \, \bar{d}_L d_R  
+   \frac{\delta_{P+S}}{v^2} \ \bar{e}_R e_L \, \bar{u}_R u_L  
+   \frac{\delta_T}{v^2} \ \bar{e}_R \sigma_{\mu \nu}  e_L \, \bar{u}_R 
\sigma^{\mu \nu} u_L   + {\rm h.c.} ~, 
\label{eq:leffNC}
\eeq
where $\delta_{P-S}=  [\hat{\alpha}_{qde}^\dagger]_{11}$, $\delta_{P+S}=  - [\hat{\alpha}_{ql}^\dagger \,   V^\dagger]_{11}$, $\delta_{T}=  - [\hat{\alpha}_{ql}^{t \ \dagger}   \, V^\dagger]_{11}$. 
These effective couplings $\delta_{S\pm P,T}$ differ from the couplings $\epsilon_{S \pm P,T}$ that contribute to beta decays at low energy and $p p \to e \nu + X$ at the LHC.  

However a simple relation arises in the generic case in which (i) all $\hat{\alpha}_{ij}$ are of the same order; (ii)  the off-diagonal entries $\hat{\alpha}_{i \neq j}$ are suppressed compared to the diagonal entries:
\beq
\delta_{P-S} = \frac{\eP - \eS}{2}  \qquad
\delta_{P+S} = \frac{\eP + \eS}{2}  \  |V_{ud}|^2 \qquad 
\delta_{T} =   \eT \,  |V_{ud}|^2  \ ~.
\label{eq:CCvsNC}
\eeq
Under these weak assumptions on the flavor structure of the new-physics couplings, we now use the Drell-Yan process to obtain bounds on $\eS \pm \eP$ and $\eT$.

To leading order in strong interactions, the NP contribution to the cross-section for  $pp \to e^+ e^- + X$ with di-lepton invariant mass $m_{ee}>\overline{m}_{ee}$ reads
\begin{subequations}
\label{eq:sigNC}
\bea
\sigma_{NP} (m_{ee}>\overline{m}_{ee})	&=&  \sigma_u \   \left[  |\delta_{P+S}|^2  + \frac{16}{3} \, |\delta_T|^2 \right] \ + \ \sigma_d   \ \,  |\delta_{P-S}|^2~, \\
\sigma_{u,d}							&=& \frac{1}{48 \pi} \frac{s}{v^4} \int_{\bar{\tau}}^{1} d \tau  \  \tau \, \tilde{L}_{u,d}  (\tau) ~,\\
\tilde{L}_q (\tau)						&=&  \int ^{-\frac{1}{2} \ln \tau}_{\frac{1}{2} \ln \tau} d y_P  \ f_{q}(\sqrt{\tau} e^{y_P}) f_{\bar{q}} (\sqrt{\tau} e^{-y_P}) ~,
\eea
\end{subequations}
where $\bar{\tau} = \overline{m}^2_{ee}/s$. The $\sigma_{u,d}(\overline{m}_{ee})$ functions are shown in Fig.~\ref{fig:sig2}, using the CTEQ6M PDF set.

Exactly in the same way as it was done for the $pp\to e\nu$ channel in the previous section, we obtain bounds on the NP couplings $\delta_\alpha$ by requiring the number of NP signal events $n_s$ to be smaller than the $90 \%$ C.L. upper bound $n_s^{up}$, i.e. $\sigma_{NP}  <  n_s^{up} / (\epsilon_{eff}  \ L)$. In order to do that, we use results from the analysis of 5 fb$^{-1}$ of dilepton data at $\sqrt{s}=7$ TeV given by the CMS Collaboration \cite{EEcms5fb}. The relevant data and details of their analysis are given in Table~\ref{tab:dataLHCee}.

\begin{table}[tbh]
\centering
\begin{tabular}{|c c c c c c c c|}
\hline
							& Reference						&	$\sqrt{s}$		&	${\cal L}$		&	$m_{ee}$	&		$n$		&	$n_b$			&		$n_s^{up}$		\\
\hline
LHC-7 		& CMS Coll. \cite{EEcms5fb}		&	7 TeV			&	5.0 fb$^{-1}$		&	1.35 TeV	&		0		&	-				&		2.3	\\
LHC-8 		& Projection						&	8 TeV			&	20 fb$^{-1}$	&	2.0 TeV		&		0		& 	$<1$			&		2.3	\\
\hline
\end{tabular}
\caption{Experimental data (real or expected) and cuts over the dilepton invariant mass used in our analyses of the $pp\to e^+e^-+X$ channel.
$n$  ($n_b$) is the number of observed  (background) events, and $n_s^{up}$ is the 
90\% CL upper limit on the number of signal events. 
\label{tab:dataLHCee}}
\end{table}

Compared to the $e+MET+X$ signature, here in the dilepton channel the geometric acceptances are lower, due to the reduced acceptance of the additional lepton compared to the MET. For the scalar and pseduoscalar interactions we find at $\sqrt{s}=7$ TeV the geometric acceptance has a mild variation from $\sim 87\%-90\%$ as the cut on $m_{ee}$ varies from 1 TeV to 1.6 TeV. The geometric acceptance of the tensor interaction is lower, and is almost constant at $\sim 75\%$ across the same mass range. 
We find that the efficiencies are essentially unchanged in moving from $\sqrt{s}=7 $ TeV to  8 TeV. Taking into account the dilepton detection efficiency of $\approx 85\%$ \cite{EEcms5fb}, we find $\epsilon_{eff}=0.75$ for the $\delta_{S\pm P}$ interactions, and $\epsilon_{eff}=0.65$ for the tensor interaction (see Table~\ref{tab:EfficienciesEE}). 

Using Eq.~\eqref{eq:CCvsNC}, we translate the bounds on $\delta_{S,P,T}$ into bounds on $\epsilon_{S,P,T}$, assuming only one of these $\epsilon$'s is non-zero at a  given time. These results are shown in Table \ref{tab:EfficienciesEE}. Also shown are estimates of future bounds that could be obtained with $20$ fb$^{-1}$ data at $\sqrt{s}=8$ TeV. To obtain these results we used Pythia 8.1 \cite{Sjostrand:2006za,Sjostrand:2007gs} to generate background $pp \rightarrow \gamma^*,Z^* \rightarrow ee+X$ events. Applying the rapidity selection cuts, a dilepton detection efficiency of $85\%$, and $m_{ee} >2.0$ TeV, we find $\sim$0.2 background events expected with 20 fb$^{-1}$ of integrated luminosity.  To obtain the projection for future bounds we therefore assumed that after all cuts no events will be detected.

The limits we obtain from current dilepton data are stronger than those from the $pp\to e+MET$ data.

\begin{table}[tbh]
\centering
\begin{tabular}{|c c c c c c| }
\hline
			&	$\epsilon_{det}$		&	$\epsilon_{geom} (\delta_{S\pm P})$	&	$\epsilon_{geom} (\delta_T)$	&	$|\epsilon_{S,P}|$ 		&   $|\epsilon_T|$ \\
\hline
LHC-7 	 	&	$\sim0.85$			& 	0.90						&	0.75				&	$1.0  \times 10^{-2}$		&	$1.3 \times 10^{-3}$	\\
LHC-8 		& 	0.85 (assumed)		& 	0.90						&	0.75				&	$0.7\times 10^{-2}$		&	$0.9 \times 10^{-3}$	\\
\hline
\end{tabular}
\caption{LHC bounds (first row) and projected bounds (second row) at 90\% C.L. for the new physics couplings $\epsilon_{S,T,P}$ at the renormalization scale $\mu = 2$~GeV in the $\overline{\rm MS}$ scheme,  obtained from the $pp\to e^+e^-+X$ channel, along with the detection efficiency and geometric acceptance.}
\label{tab:EfficienciesEE}
\end{table}

\subsection{Discussion of LHC bounds}

We can summarize the results  of this section as follows:

(i) We have updated the bounds on $\epsilon_{S,T}$ obtained from the $pp\to e +MET+X$ channel using 5 fb$^{-1}$ of data (see Table \ref{tab:resultsLHCenu}), improving by $\sim 25\%$ the bounds obtained with 1 fb$^{-1}$ in Ref.~\cite{Bhattacharya:2011qm}. 

(ii)  We have extended this analysis to extract  bounds on the other eight  effective 
 couplings $\epsilon_i$ and $\tilde{\epsilon}_i$ that describe non-standard CC weak interactions. 
We show that for $\epsilon_{P,S,T}$, $\tilde{\epsilon}_{P,S,T,R}$ and $\eLc$ the current bounds from an analysis of the $pp\to e+MET + X$
channel at the LHC are at the  0.1\%-1\%  level (see Table \ref{tab:resultsLHCenu}), whereas this search is essentially insensitive to the rest of couplings: $\eR$, $\eLv$ and $\teL$. 

(iii) We have studied the impact of   $pp\to e^+ e^- + X$ in constraining the couplings $\epsilon_{S,P,T}$. 
This analysis relies on the use of $SU(2)$ gauge invariance and very  weak assumptions about the flavor structure of the theory.  
The resulting bounds on  $\epsilon_{S,P,T}$ 
are  stronger  than those obtained in the $pp\to e +MET+X $ channel by a factor 1.3 and 2.3 for $\epsilon_{S,P}$ and $\eT$ respectively 
(compare Tables   \ref{tab:resultsLHCenu} and  \ref{tab:EfficienciesEE}).

\section{Conclusions}
\label{sect:conclusions}

In this work we have analyzed all the  phenomenological handles on 
the effective  couplings $\epsilon_i$ and $\tilde{\epsilon}_i$ 
characterizing non-standard CC interactions.
We have reviewed the limits on  $\epsilon_i$ and $\tilde{\epsilon}_i$ 
from low energy experiments (nuclear and neutron $\beta$ decays, CKM unitarity tests, pion decays, ...), 
and we have computed  the bounds on the same couplings 
arising  from LHC searches.  
This direct comparison relies on the assumption that the new physics mediating these new interactions 
arises at very high scale $\Lambda > $ few TeV. 
Comparison of the best bounds available for each interaction from low- and high-energy 
experiments are shown in   Tables \ref{tab:summaryL} and  \ref{tab:summaryR}, for the non-standard  couplings involving left-handed and 
right-handed neutrinos, respectively. 
The main points can be summarized as follows:
\begin{itemize}

\item For the pseudo-scalar couplings $\eP$ and $\teP$ the low-energy constraints from pion decay are at the $10^{-4}$ level, 
very hard to reach at the LHC in the near future.  The same applies to the vector interactions $\eLc$, $\eLv$ and $\eR$, 
for which the 90\% C.L. bound from CKM-unitarity (and LEP physics) is $5 \times 10^{-4}$;

\item For scalar and tensor interactions with left-handed neutrinos $\eS$ and $\eT$,  the low-energy experiments 
 have been traditionally stronger, but the current LHC bounds have caught up and both probes are at the $10^{-2}$ and $10^{-3}$ level for $\eS$ and $\eT$ respectively. 
In the next few years we expect improvements in the bounds from both the LHC 
and  low-energy, through  neutron~\cite{WilburnUCNB,abBA,Pocanic:2008pu,UCNb}  
and $^6$He decay~\cite{Knecht201143} measurements at the 0.1\% level. 
These considerations imply that low-energy searches with $10^{-4}$ sensitivity would have 
unmatched constraining potential, even in the LHC era.

\item For scalar and tensor interactions with right-handed neutrinos $\teS$ and $\teT$, the LHC bounds are also at the $10^{-2}$ and $10^{-3}$ level respectively, significantly better than current and future 
 low-energy limits (to match the LHC bound one needs measurements of the 
 electron-neutrino correlation ``$a$"  in Gamow Teller transitions at the level of $\delta a_{GT}/a_{GT} \sim 0.01$\%) 
 \footnote{Within a given NP model the ratio $\Gamma(\pi\to e\nu) / \Gamma(\pi\to \mu\nu)$ is likely to produce the strongest bound not only on
$\epsilon_P$ and $\teP$,  but also on $\epsilon_{S,T}$ and $\tilde{\epsilon}_{S,T}$,  through their loop-induced contribution. However, as explained in Section~\ref{sect:low-energy} and Ref. \cite{Bhattacharya:2011qm}, there is a loophole to this argument, because cancelations between different contributions can occur. For this reason in a general model-independent analysis the LHC provides the strongest constraint on $\tilde{\epsilon}_{S,T}$.}.  
And the same conclusion applies to $\teR$,  for which the LHC bound is $5\times 10^{-3}$ and no significant limit is available from low-energy probes;

\item The remaining coefficient $\teL$ cannot be probed at such a strong level by any of the mentioned experiments.   

\end{itemize}

In conclusion, we have shown that by combining low-energy and LHC searches, 
a more complete  and accurate picture of non-standard CC interactions emerges.  
As we discussed earlier,  such a  direct comparison 
is only possible if  the new physics  originates in the multi-TeV scale. 
The next natural question is: how robust are the above  conclusions 
when  the mediators of the new interactions  
are not heavy enough to be integrated out at LHC energies?  
In this case the EFT treatment of LHC data breaks down and 
the discussion becomes necessarily model-dependent. 
We leave investigations of  explicit classes of models for future work.

\begin{table}
\centering
\begin{tabular}{|c|c|c|c|c|c|c|}
\hline
                  & $ |\epsilon_L^{(v)}| $           &  	 $ \epsilon_L^{(c)}$      &       $|\epsilon_R|$           &     $|\epsilon_P |$    &      $|\epsilon_S|$         &       $| \epsilon_T|$    \\
\hline 
\hline
Low energy	&   		0.05	                   &          0.05                    &               0.05                           &            0.06             &                 0.8                 &              0.1                 \\
\hline 
 LHC		$(e \nu)$   &   		-	                 &	        (-0.3,+0.8)	       &               -                   &               1.3                   &                1.3                       &             0.3               \\
\hline 
  LHC 
$(e^+ e^-)$  &   		-		          	&		-	            &                 -                     &           1.0        &               1.0               &          0.1                  \\
\hline
\end{tabular} 
\caption{Summary of 90\% CL bounds (in units of  $10^{-2}$)  
on the non-standard couplings $\epsilon_\alpha$  obtained  from low-energy and LHC searches. 
In order to deduce the low-energy  bound  on the scalar coupling 
we used  $g_S = 0.8(4)$~\cite{Bhattacharya:2011qm}.  
Using  $g_S = 1.08(28)$~\cite{Green:2012ej} would lead to the stronger bound $|\epsilon_S| < 0.4 \times 10^{-2}$. 
\label{tab:summaryL}}
\end{table}

\begin{table}
\centering
\begin{tabular}{|c|c|c|c|c|c|}
\hline
				&	$ |  \teL| $				&	$| \teR| $           				&	$| \teP|$			&	$| \teS|$			&	$|\teT|$               \\
\hline 
\hline
Low energy	&		6					&		6						&	0.03				&		14			&	3.0				\\
\hline 
LHC $(e \nu)$		&		-					&	0.5							&	1.3				&		1.3			&	0.3				\\
\hline
\end{tabular} 
\caption{Summary of 90\% CL bounds (in units of  $10^{-2}$)  
on the non-standard couplings $\tilde{\epsilon}_\alpha$  obtained  from low-energy and LHC searches.
In order to deduce the low-energy bounds on the scalar coupling 
we used  $g_S = 0.8(4)$ and $g_T = 1.05(35)$~\cite{Bhattacharya:2011qm}.
Using  $g_S = 1.08(28)$~\cite{Green:2012ej} would lead to the stronger bound $|\tilde{\epsilon}_S| < 6.9 \times 10^{-2}$. 
\label{tab:summaryR}}
\end{table}

\section*{Acknowledgements} 
We thank  Tanmoy Bhattacharya and  Rajan Gupta for discussions, and Bruce Campbell for correspondence concerning  Ref.~\cite{Campbell:2003ir}. MGA thanks  the LANL T-2 Group for its hospitality and support during the completion of this work. VC and MG acknowledge  support by the DOE  Office of Science  and the LDRD program at Los Alamos  National Laboratory. MGA was supported by the U.S. DOE contract DE-FG02-08ER41531 and by the Wisconsin Alumni Research Foundation.


\bibliographystyle{h-physrev}
\bibliography{Vincenzo,Michael,Martin}

\begin{thebibliography}{10}

\bibitem{Weinberg:2009zz}
S.~Weinberg,
\newblock J.Phys.Conf.Ser. {\bf 196}, 012002 (2009).

\bibitem{Severijns2006dr}
N.~Severijns, M.~Beck, and O.~Naviliat-Cuncic,
\newblock Rev. Mod. Phys. {\bf 78}, 991 (2006), nucl-ex/0605029.

\bibitem{Herczeg2001vk}
P.~Herczeg,
\newblock Prog. Part. Nucl. Phys. {\bf 46}, 413 (2001).

\bibitem{Profumo:2006yu}
S.~Profumo, M.~J. Ramsey-Musolf, and S.~Tulin,
\newblock Phys.Rev. {\bf D75}, 075017 (2007), hep-ph/0608064.

\bibitem{Bauman:2012fx}
S.~Bauman, J.~Erler, and M.~Ramsey-Musolf,
\newblock (2012), 1204.0035.

\bibitem{Lee:1956qn}
T.~Lee and C.-N. Yang,
\newblock Phys.Rev. {\bf 104}, 254 (1956).

\bibitem{Jackson1957zz}
J.~D. Jackson, S.~B. Treiman, and H.~W. Wyld,
\newblock Phys. Rev. {\bf 106}, 517 (1957).

\bibitem{Bhattacharya:2011qm}
T.~Bhattacharya {\em et~al.},
\newblock Phys.Rev. {\bf D85}, 054512 (2012), 1110.6448.

\bibitem{Nakamura:2010zzi}
Particle Data Group, K.~Nakamura {\em et~al.},
\newblock J.Phys.G {\bf G37}, 075021 (2010).

\bibitem{Green:2012ej}
J.~Green {\em et~al.},
\newblock (2012), 1206.4527.

\bibitem{Cirigliano:2009wk}
V.~Cirigliano, J.~Jenkins, and M.~Gonzalez-Alonso,
\newblock Nucl.Phys. {\bf B830}, 95 (2010), 0908.1754.

\bibitem{Charles:2004jd}
CKMfitter Group, J.~Charles {\em et~al.},
\newblock Eur. Phys. J. {\bf C41}, 1 (2005), hep-ph/0406184.

\bibitem{Mateu:2007tr}
V.~Mateu and J.~Portoles,
\newblock Eur.Phys.J. {\bf C52}, 325 (2007), 0706.1039.

\bibitem{Britton:1992pg}
D.~Britton {\em et~al.},
\newblock Phys.Rev.Lett. {\bf 68}, 3000 (1992).

\bibitem{Czapek:1993kc}
G.~Czapek {\em et~al.},
\newblock Phys.Rev.Lett. {\bf 70}, 17 (1993).

\bibitem{Cirigliano:2007xi}
V.~Cirigliano and I.~Rosell,
\newblock Phys.Rev.Lett. {\bf 99}, 231801 (2007), 0707.3439.

\bibitem{Hardy:2008gy}
J.~Hardy and I.~Towner,
\newblock Phys.Rev. {\bf C79}, 055502 (2009), 0812.1202.

\bibitem{Bychkov:2008ws}
M.~Bychkov {\em et~al.},
\newblock Phys.Rev.Lett. {\bf 103}, 051802 (2009), 0804.1815.

\bibitem{Voloshin:1992sn}
M.~Voloshin,
\newblock Phys.Lett. {\bf B283}, 120 (1992).

\bibitem{Herczeg:1994ur}
P.~Herczeg,
\newblock Phys.Rev. {\bf D49}, 247 (1994).

\bibitem{Campbell:2003ir}
B.~A. Campbell and D.~W. Maybury,
\newblock Nucl.Phys. {\bf B709}, 419 (2005), hep-ph/0303046.

\bibitem{WilburnUCNB}
W.~Wilburn {\em et~al.},
\newblock Rev. Mex. Fis. {\bf Suppl. 55}, 119 (2009).

\bibitem{abBA}
R.~Alarcon {\em et~al.},
\newblock {Precise Measurement of Neutron Decay Parameters}, 2007.

\bibitem{Pocanic:2008pu}
Nab Collaboration, D.~Pocanic {\em et~al.},
\newblock Nucl.Instrum.Meth. {\bf A611}, 211 (2009), 0810.0251.

\bibitem{UCNb}
K.~P. Hickerson,
\newblock {The Fierz Interference Term in Beta-Decay Spectrum of UCN}, 2009,
\newblock UCN Workshop, November 6--7 2009, Santa Fe, New Mexico.

\bibitem{Knecht201143}
A.~Knecht {\em et~al.},
\newblock Nucl.Instrum.Meth. {\bf A660}, 43 (2011).

\bibitem{Jackson1957206}
J.~D. Jackson, S.~B. Treiman, and H.~W. Wyld,
\newblock Nuclear Physics {\bf 4}, 206  (1957).

\bibitem{Barger:1983wf}
V.~D. Barger, A.~D. Martin, and R.~Phillips,
\newblock Z.Phys. {\bf C21}, 99 (1983).

\bibitem{Pumplin:2002vw}
J.~Pumplin {\em et~al.},
\newblock JHEP {\bf 0207}, 012 (2002), hep-ph/0201195.

\bibitem{ENcms5fb}
CMS Collaboration, S.~Chatrchyan {\em et~al.},
\newblock JHEP {\bf 1208}, 023 (2012), 1204.4764.

\bibitem{Martin:2009iq}
A.~D. Martin, W.~J. Stirling, R.~S. Thorne, and G.~Watt,
\newblock Eur. Phys. J. {\bf C63}, 189 (2009), 0901.0002.

\bibitem{ATLAS:2012dm}
ATLAS Collaboration, G.~Aad {\em et~al.},
\newblock (2012), 1209.4446.

\bibitem{CMS8TeV}
{CMS Collaboration, Search for leptonic decays of W' bosons in pp collisions at
  sqrt(s)=8 TeV, CMS-PAS-EXO-12-010}.

\bibitem{Broadhurst:1994se}
D.~J. Broadhurst and A.~Grozin,
\newblock Phys.Rev. {\bf D52}, 4082 (1995), hep-ph/9410240.

\bibitem{EEcms5fb}
CMS Collaboration, S.~Chatrchyan {\em et~al.},
\newblock (2012), 1206.1849.

\bibitem{Sjostrand:2006za}
T.~Sjostrand, S.~Mrenna, and P.~Z. Skands,
\newblock JHEP {\bf 0605}, 026 (2006), hep-ph/0603175.

\bibitem{Sjostrand:2007gs}
T.~Sjostrand, S.~Mrenna, and P.~Z. Skands,
\newblock Comput.Phys.Commun. {\bf 178}, 852 (2008), 0710.3820.

\end{thebibliography}

\end{document}